\documentstyle[preprint,aps]{revtex}
\input epsf.sty
\tighten

\begin{document}
\draft

\title{\Large \bf Comment on "Universal Fluctuations
                       in Correlated Systems"}

\author{\bf B. Zheng and S. Trimper$^2$}

\address{$^1$ Physics Department, Zhejiang University, 
              Hangzhou 310027, P.R. China}
\address{$^2$ FB Physik, Universit\"at Halle, 06099 Halle, Germany}
\address{$^3$ICTP, 34014 Trieste, Italy}

\maketitle


In \cite {bram00}, it is suggested that the probability distribution
functions (PDF's) of fluctuating observables for critical systems
in different universality classes exhibit a same form.
This Comment concerns the PDF's for equilibrium systems.

\begin{figure}[t]\centering 
\epsfysize=10.cm 
\epsfclipoff 
\fboxsep=0pt
\setlength{\unitlength}{1cm} 
\begin{picture}(13.6,12.)(0,0)
\put(-1.,0){{\epsffile{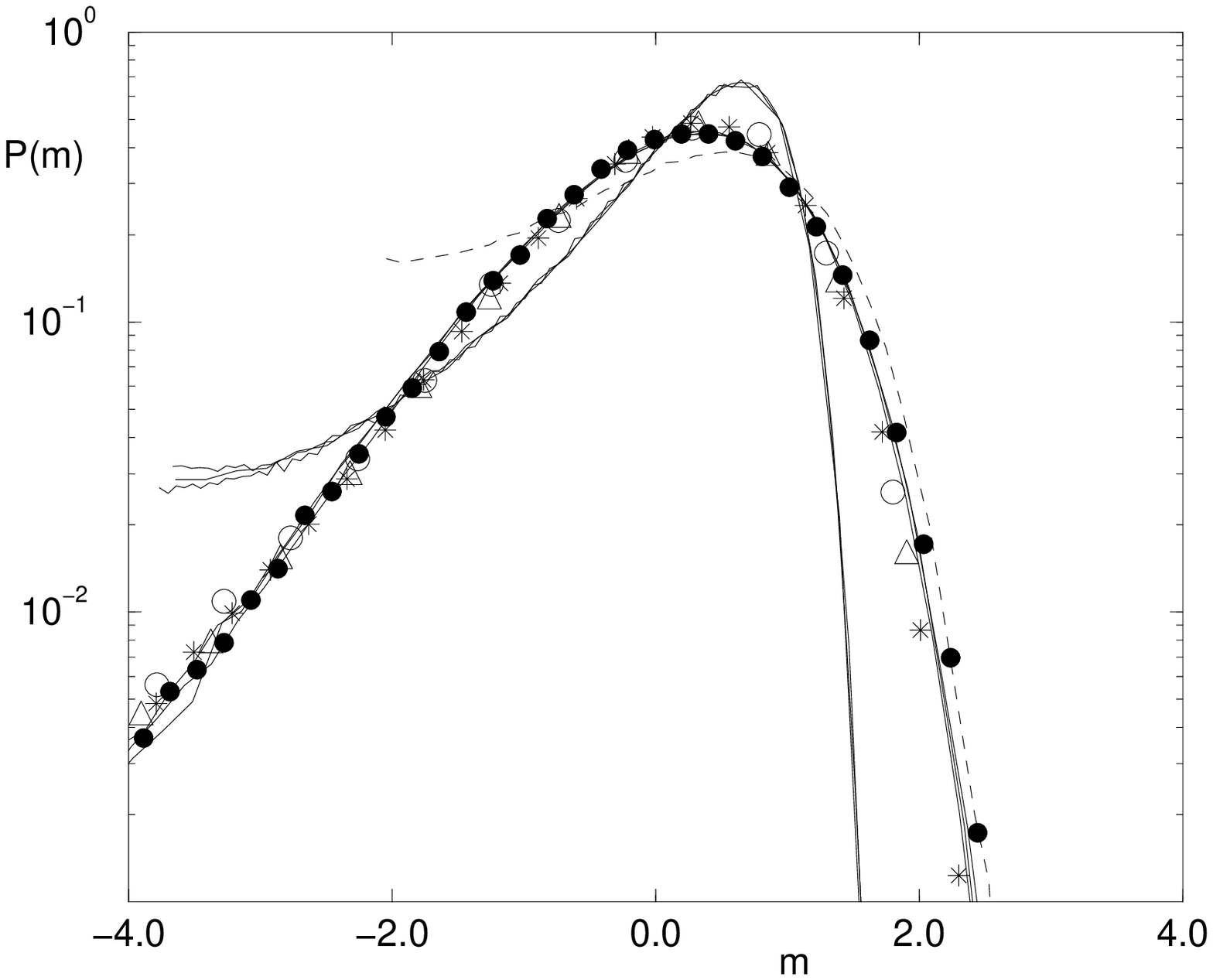}}} 
\end{picture} 
\caption{The three solid lines with a higher peak are
 for the 2D Ising model at $T_c$ with $L=32$, $64$ 
 and $128$. The three solid lines with a lower peak
 are for the 2D XY model at $T=0.89$ with $L=16$, $32$ 
 and $64$. The dashed curve is for the 3D Ising model
 at $T_c$ with $L=32$. The circles, triangles and
 stars are for the 2D Ising model at $s=2.90$ with $L=32$, $64$ 
 and $128$. The filled circles are for the 3D Ising 
 model at $s=2.21$ with $L=32$.
} 
\label{f1}
\end{figure}

In \cite {aji01}, it is pointed out that
the standard scaling form is a sufficient condition
for the data collapse. The suggestion in \cite {bram00}
seems contradicting the idea of standard universality.
 To clarify it,
we have performed calculations for different critical systems. 
Firstly, the temperature is fixed at the bulk critical
value $T_c$. In Fig. \ref {f1},
the normalized PDF $P(m)$ 
for the Ising and XY models are displayed. Here
$m=(M-<M>)/\sigma$ with $M$ being the magnetization,  
$<M>$ being its mean and $\sigma$ being the standard 
deviation. Data collapse for different lattice sizes
$L$'s are observed.
However, the differences between the curves 
for different models are clearly
not minor perturbations. The PDF for the Potts model
is even not symmetric in $M$ and can hardly be put
in the figure.

The critical temperature can be modified in a finite system.
Let us take a size-dependent coupling $K(L) \sim 1/T(L)$
such that $\tau \equiv (K(L)-K_c)/K_c=s / L^{1/\nu}$ 
with $\nu=1$ being
the static exponent and $s$ being a constant. 
Following similar scaling analysis in \cite {aji01}
(see also \cite {bram01}), it leads to 
$P(m,\tau,L)=P(m,L^{1/\nu} \tau)=P(m,s)$.
At the critical regime, data for different $L$'s at $T(L)$'s also
collapse onto a single curve, 
 but $P(m,s)$ changes continuously with $s$.
 If $s$ is small, in other words, 
 if the spatial correlation function $l(L)$ is much larger
 than the lattice size $L$, $T(L)$ can be considered
 approximately as a size-dependent critical temperature.
Choosing $s=2.90$, $P(m,s)$ looks
falling onto that for the 2D XY model at $T=0.89$.
This is shown in Fig. \ref {f1}. 
However, $T(L)$ with such a large value of $s$ 
should not be defined as 
a size-dependent critical temperature, since in the infinite limit
of $L$, the behavior of the system (not only $P(m)$)
remains very different from
that at the bulk $T_c$. To confirm this, we have calculated
the spatial correlation function and found that 
at $s=2.90$
the correlation length $l(L)$ 
is much smaller than the lattice size $L$. 

We do not think the {\it shape} 
of $P(m)$ for the XY model is 
a characteristic property 
at (or very close to) the critical point.
 The observation is that for systems with a second
order transition such as the Ising and Potts models,
the tail of the PDF for negative $m$ at the $T_c$
reaches a nonzero value at $M=0$ and is
roughly power-law-like (before cut at $M=0$).
This reasonably indicates that the system
can transit from the positive sector of $M$ 
to the negative one.
 Below $T_c$, symmetry breaking occurs.
The (infinite) system can not transit from one sector to another.
The exponential-like tail at a large but not too large $s$
should be a signal of symmetry breaking.
(But when $s$ tends to infinite, the tail
crosses over to Gaussian.)
For the XY model,
the fluctuations are mainly rotational. The exponential tail 
of $P(m)$ is only an indication
for the energy barrier in small $M$ regime.
The conjecture is that the exponential-like
tail of the PDF induced typically by energy barriers
may be similar to the exponential decay 
of the correlation functions
with a finite correlation length.
It can be rather generic, and independent of whether
the system is with or without a first order, second order 
or Kosterlitz-Thouless phase transition. 
The results for the 1D and 3D XY model in \cite {bram01}
support this statement.
A critical point is only a sufficient condition
for data collapse for infinite systems.

The complete form of the PDF is in general 
not independent of universality classes.
However, the large $s$ regime may be somewhat special
and it needs more investigations. For example,
as shown in Fig. \ref {f1}, $P(m,s)$ for the 3D Ising model
at $s=2.21$ fits also to the curve of the XY model
at $T=0.89$. 

Acknowledgement: Work is supported in part by
DFG, TR 300/3-1.


\end{document}